\documentclass[prl,aps,longbibliography,amssymb,twocolumn,superscriptaddress]{revtex4-1}
\usepackage{amsmath}
\usepackage{amssymb}
\usepackage{amsthm}
\usepackage{amsfonts}
\usepackage{listings}
\usepackage{enumerate}
\usepackage{latexsym}
\usepackage{psfrag}
\usepackage{bm}
\usepackage[all]{xy}
\usepackage{graphicx}
\usepackage{subfigure}
\usepackage[pdftex,colorlinks]{hyperref}
\usepackage{color}
\usepackage{mathtools}
\usepackage{eufrak}

\begin{document}

\title{Raman evidence for dimerization and Mott collapse in $\alpha$-RuCl$_3$ under pressures}

\author{Gaomin Li}
\affiliation{School of Advanced Materials, Shenzhen Graduate School Peking
  University, Shenzhen 518055, P. R. China}
\affiliation{Shenzhen Institute for Quantum Science and Engineering, and Department of Physics, Southern University of Science and Technology, Shenzhen 518055, China}
\author{Xiaobin Chen}
\affiliation{School of Science, Harbin Institute of Technology, Shenzhen 518055, China}
\author{Yuan Gan}
\author{Fenglei Li}
\author{Mingqi Yan}
\affiliation{Department of Physics, Southern University of Science and Technology, Shenzhen 518055,
  China}
\author{Shenghai Pei}
\author{Yujun Zhang}
\affiliation{Shenzhen Institute for Quantum Science and Engineering, and Department of Physics, Southern University of Science and Technology, Shenzhen 518055, China}
\author{Le Wang}
\affiliation{Institute of Physics, Chinese Academy of Sciences, Beijing 100190, China}
\author{Huimin Su}
\author{Junfeng Dai}
\affiliation{Department of Physics, Southern University of Science and Technology, Shenzhen 518055,
  China}
\author{Yuanzhen Chen}
\affiliation{Shenzhen Institute for Quantum Science and Engineering, and Department of Physics, Southern University of Science and Technology, Shenzhen 518055, China}
\affiliation{Shenzhen Key Laboratory of Quantum Science and Engineering, Shenzhen 518055, PR China.}

\author{Youguo Shi}
\affiliation{Institute of Physics, Chinese Academy of Sciences, Beijing 100190, China}

\author{XinWei Wang}
\affiliation{School of Advanced Materials, Shenzhen Graduate School Peking
  University, Shenzhen 518055, P. R. China}
\author{Liyuan Zhang}
\author{Shanmin Wang}
\affiliation{Department of Physics, Southern University of Science and Technology, Shenzhen 518055,
  China}
\author{Dapeng Yu}
\affiliation{Shenzhen Institute for Quantum Science and Engineering, and Department of Physics, Southern University of Science and Technology, Shenzhen 518055, China}
\affiliation{Shenzhen Key Laboratory of Quantum Science and Engineering, Shenzhen 518055, PR China.}

\author{Fei Ye}
\email{yef@sustc.edu.cn}
\author{Jia-Wei Mei}
\email{meijw@sustc.edu.cn}
\affiliation{Shenzhen Institute for Quantum Science and Engineering, and Department of Physics, Southern University of Science and Technology, Shenzhen 518055, China}
\author{Mingyuan Huang}
\email{huangmy@sustc.edu.cn}
\affiliation{Shenzhen Institute for Quantum Science and Engineering, and Department of Physics, Southern University of Science and Technology, Shenzhen 518055, China}
\affiliation{Shenzhen Key Laboratory of Quantum Science and Engineering, Shenzhen 518055, PR China.}

\date{\today}

\begin{abstract}
  We perform Raman spectroscopy studies on $\alpha$-RuCl$_3$ at room temperature to explore its phase transitions of magnetism and chemical bonding under pressures. The Raman measurements resolve two critical pressures, about $p_1=1.1$~GPa and $p_2=1.7$~GPa, involving very different intertwining behaviors between the structural and magnetic excitations. With increasing pressures, a stacking order  phase transition of $\alpha$-RuCl$_3$ layers develops at $p_1=1.1$~GPa, indicated by the new Raman phonon modes and the modest Raman magnetic susceptibility adjustment.
  The abnormal softening and splitting of the Ru in-plane Raman mode provide direct evidence of the in-plane dimerization of the Ru-Ru bonds at $p_2=1.7$~GPa. The Raman susceptibility is greatly enhanced with pressure increasing and sharply suppressed after the dimerization. We propose that the system undergoes Mott collapse at $p_2=1.7$~GPa and turns into a dimerized correlated band insulator. Our studies demonstrate competitions between Kitaev physics, magnetism, and
chemical bondings in Kitaev compounds.
\end{abstract}

\maketitle

\textit{Introduction--} The spin-orbit coupling always invigorates new vitality to the intertwines of magnetism and chemical bonds~\cite{Goodenough1963,Dzyaloshinsky1958,Moriya1960,Shekhtman1992,Pesin2010}, and generates the bond-dependent
Dzyaloshinsky-Moriya-type~\cite{Dzyaloshinsky1958,Moriya1960} and Ising-type interactions~\cite{Shekhtman1992}. While the former interactions
yield non-trivial
magnetic topology \cite{Nagaosa2012,Nagaosa2012a}, the latter terms on a honeycomb
lattice provide a pathway to realize the Kitaev exactly solvable
spin model~\cite{Kitaev2006, Jackeli2009,Chaloupka2010}. The ground state of
Kitaev spin model~\cite{Kitaev2006} represents a typical quantum paramagnetism
dubbed  quantum spin liquid, in which the spin degree of
freedom does not freeze even at zero temperature~\cite{Anderson1987}. Quantum
spin liquid displays various patterns of long-range
quantum entanglement~\cite{Wen2004,Kitaev2006,Levin2006} and supports the fractional
excitations~\cite{Laughlin1983,Kivelson1987,Read1989,Read1991,Wen1991,Ye2015,Ye2017}.
Kitaev interactions have been
identified in layered honeycomb magnetic materials, such as
$\alpha$-Li$_2$IrO$_3$\cite{Biffin2014} and $\alpha$-RuCl$_3$\cite{Plumb2014}.
However, due to further non-Kitaev interactions, these materials
have long-range magnetic orders at low
temperatures~\cite{Johnson2015,Sears2015,Banerjee2016,Williams2016,Ran2017}. To achieve the quantum spin liquid state,   in-plane magnetic
fields have been implemented to suppress the magnetic order in $\alpha$-RuCl$_3$, and the magnetic properties are consistent with theoretical expectations~\cite{Zheng2017,Yu2018,Jansa2018,Baek2017,Banerjee2018,Wolter2017,Wang2017a,Hentrich2018,Kasahara2018a}.

Pressure also promotes the breakdown of magnetic order in $\alpha$-RuCl$_3$~\cite{Wang2017,Cui2017,Bastien2018,Biesner2018}.   Above a critical pressure, the magnetic
signal disappears~\cite {Wang2017,Cui2017,Bastien2018}; however, the charge gap
does not change significantly, and the system remains an insulating
state~\cite{Wang2017}. A present debate is whether the phase transition under
pressures involves the structural deformation due to chemical bondings. X-ray
diffraction (XRD) measurements in Ref.~\cite{Wang2017} didn't detect any crystal
structural phase transition up to 150 GPa, and hence a new quantum magnetic
disordered state was proposed. However, XRD measurements in Ref.~\cite{Bastien2018}
revealed a Ru-Ru bond dimerization of about 0.6~\AA ~above a critical pressure,
and supported a non-magnetic gapped dimerized state at high pressures. Ref.~\cite{Biesner2018}
reached a similar non-magnetic dimerized scenario in optical measurements.
This is reminiscent of recent high-pressure investigations on
the 2D Kitaev material $\alpha$-Li$_2$IrO$_3$~\cite{Hermann2018} and its 3D
polymorph $\beta$-Li$_2$IrO$_3$~\cite{Majumder2018}. At high pressures,
$\alpha$-Li$_2$IrO$_3$ dimerizes~\cite{Hermann2018}, while
$\beta$-Li$_2$IrO$_3$ manifests the coexistence of dynamically correlated and frozen spins
without structural deformation~\cite{Majumder2018}. Raman spectrum simultaneously detects the lattice and magnetic
excitations, and their mutual couplings~\cite{Lemmens2003,Devereaux2007}. It is an exemplary experimental
tool to study competition between spin-orbit couplings~\cite{Moriya1968,Fleury1968,Nasu2016,Fu2017}, magnetism~\cite{Shastry1990,Lemmens2003,Devereaux2007,Ko2010}, and
chemical bondings in Kitaev
compounds\cite{Knolle2014a,Sandilands2015,Glamazda2016,Nasu2016,Glamazda2017}.

In this Letter, we perform Raman scattering measurements on
$\alpha$-RuCl$_3$  to study the nature of the phase transitions under pressures.
Measurements are carried out at room temperature if the temperature is
not specified. From the evolution of the Raman spectra, we
identify two characteristic pressures $p_1=1.1$~GPa and $p_2=1.7$~GPa for
structural phase transitions. The inversion symmetry of the monoclinic $C2/m$
breaks at $p_1=1.1$~GPa and the system turns out to be trigonal $P3_112$
owing to the stacking pattern changes of the $\alpha$-RuCl$_3$ layers. At
$p_2=1.7$~GPa, the Ru in-plane Raman mode (161~cm$^{-1}$ at ambient pressure)
softens and splits, indicating the dimerization of the Ru-Ru bonds. 
The Raman susceptibility is greatly enhanced with pressure increasing, mildly adjusts at $p_1=1.1$~GPa, and sharply suppressed after the dimerization at $p_2=1.7$~GPa. We conclude that the system undergoes the Mott collapse
and turns out to be dimerized correlated band insulator with the pressure larger than $p_2=1.7$~GPa.

\textit{Experimental setup--}
High-quality single crystals of $\alpha$-RuCl$_3$ are grown from commercial
RuCl$_3$ powder by chemical vapor transport method. We use the diamond anvil
cell to apply hydrostatic pressures on the samples, and calibrate the value of
pressures by the shift of the photoluminescence of Ruby. Raman spectrum
measurement is conducted in the backscattering configuration with the light
polarized in the basal plane. Light from a 633~nm and 488~nm laser is focused
down to 3 $\mu$m with the power below 1~mW. Two ultra-narrow band notch filters
are used to suppress the Rayleigh scattering light. The scattering light is
dispersed by a Horiba iHR550 spectrometer and detected by a liquid nitrogen
cooled CCD detector.

\begin{figure}[t]
\centering
\includegraphics[width=\columnwidth]{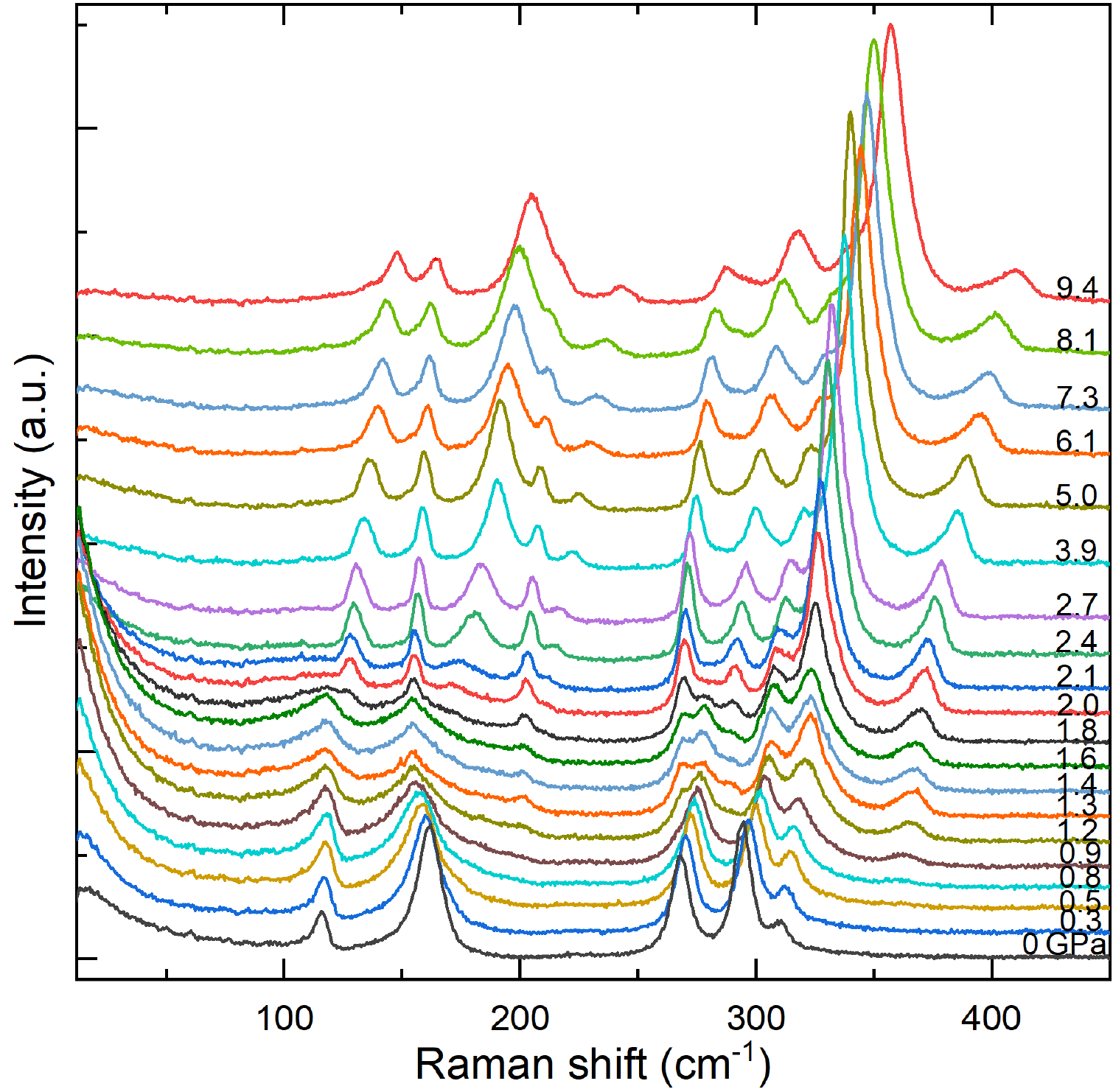}
\caption{Evolution of the Raman spectra of the  $\alpha$-RuCl$_3$ under
  different pressure at room temperature.}
\label{fig:figure1}
\end{figure}

\textit{Raman spectral evolution --}
Figure.~\ref{fig:figure1} displays the evolution of the Raman spectra of $\alpha$-RuCl$_3$ with the pressure from ambient
pressure to 9.4~GPa. The highest measured pressure is up to 24~GPa and the pressure
process is reversible~\footnote{See Supplemental Materials for more details.}. Here, we can identify two characteristic pressures, $p_1=1.1$~GPa and $p_2=1.7$~GPa, at which the dramatic change of Raman spectra implies the structural phase transitions.  At ambient pressure, five Raman modes
are clearly resolved at 116, 161, 268, 294, and 310~cm$^{-1}$. At $p_1=1.1$~GPa, three new Raman modes
at 201, 290 and 363 cm$^{-1}$ appear. The original five
modes evolves as following. For the mode at 116~cm$^{-1}$, a new mode splits out at the right side and the
original mode disappears at $p_2=1.7$~GPa. A similar splitting can be
identified for the mode at 161 cm$^{-1}$ at the same pressure, but the original mode
remains after that.  A splitting can be seen at the
left side of the mode at 268~cm$^{-1}$ at $p_1=1.1$~GPa and the original peak disappears at
$p_2=1.7$~GPa. No splitting behavior can be resolved for the mode at 294
cm$^{-1}$ and the intensity of the mode at 310 cm$^{-1}$ experiences a dramatic increasing
after 1.7~GPa. After $p_2=1.7$~GPa, no sudden change is observed up to 24 GPa~\cite{Note1}.

\begin{figure}[b]
\centering
\includegraphics[width=\columnwidth]{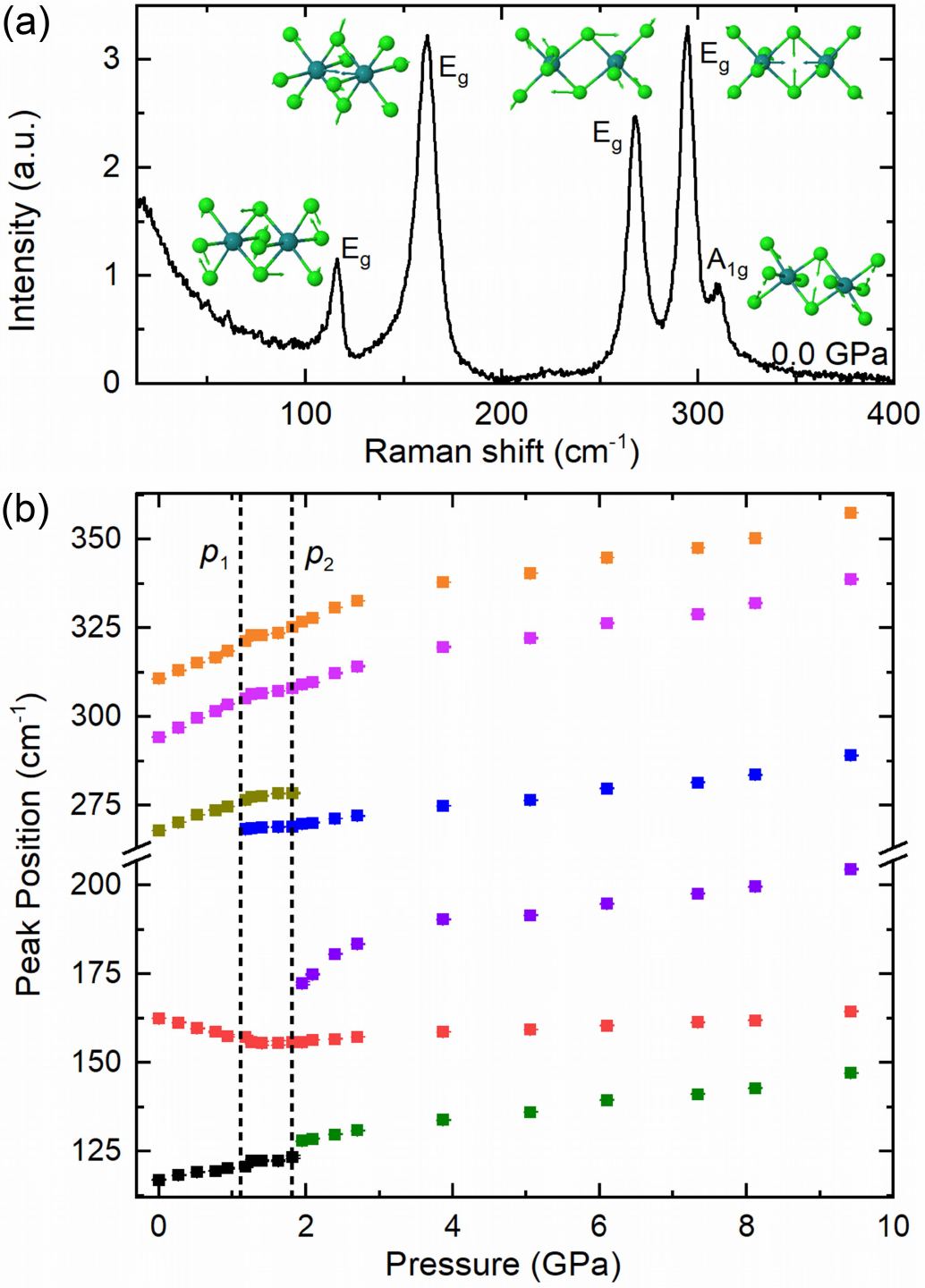}
\caption{(a) The atomic displacements of Raman mode eigenvectors for 5 Raman active
  modes ($4E_g+1A_{1g}$) in $\alpha$-RuCl$_3$ under ambient pressure. Only one represented mode is shown for the double degenerated
$E_g$ modes. (b) Pressure dependence of the frequencies of the 5 dominate Raman peaks.
  $p_1=1.1$~GPa and $p_2=1.7$~GPa are two critical pressures.  }
\label{fig:figure2}
\end{figure}
\textit{Raman phonon mode assignment --}
 At ambient pressure, we perform Raman
and IR measurements on exfoliated $\alpha$-RuCl$_3$ samples down to three atomic
layers and no significant difference is observed~\cite{Note1}. Hence we can assign the
Raman modes at ambient pressure according to the $D_{3d}$ point group of the single
$\alpha$-RuCl$_3$ layer. From group theory, the irreducible representation of
atomic displacement at the $\Gamma$ point is $\Gamma_{\text{opt}} = 2A_{1g} + 2A_{2g} + 4E_g + A_{1u} + 2A_{2u}
+ 3E_u$. Among them, Raman active modes are $\Gamma_R = 2A_{1g} + 4E_g$. From the polarization
measurement~\cite{Note1} and previous Raman studies~\cite{Sandilands2015}, the
first four modes are assigned as doubly degenerated $E_g$ mode and the last mode
as $A_{1g}$ mode. Other two small Raman modes at 219 and 339~cm$^{-1}$, can also be
resolved by using 488~nm laser as the excitation light~\cite{Note1}.


We assign the Raman mode eigenvectors with the help of first-principle
calculations~\cite{Note1}  and the atomic displacement of 5 Raman active modes are displayed in
Fig.~\ref{fig:figure2} (a). For the 4 $E_g$ modes, the mode at 116 cm$^{-1}$ is dominated by the twist
of the Ru-Cl-Ru-Cl plane; the mode at 161 cm$^{-1}$ is associated with Ru in-plane
relative movement; the mode at 268 cm$^{-1}$ is related to the Ru-Cl-Ru-Cl plane
shearing, and the mode at 294 cm$^{-1}$ is the Ru-Cl-Ru-Cl ring breathing mode. The
$A_{1g}$ mode at 310 cm$^{-1}$ can be assigned as the symmetrical layer breathing mode.
The other A$_{1g}$ mode is the triangular distortion mode and
the calculated frequency is about 149 cm$^{-1}$, which is close to the frequencies
of the $A_{1g}$ modes observed in CrCl$_3$~\cite{Glamazda2017} and FeCl$_3$~\cite{Caswell1978}, 142 and 165 cm$^{-1}$, respectively. Since there is
no A$_{1g}$ peak observed in this range, we believe that the triangular distortion
mode is unresolvable due to the small scattering cross section, other than the
small Raman mode observed at 339 cm$^{-1}$.


\textit{Inversion symmetry breaking at $p_1=1.1$~GPa --}
The
main feature at $p_1=1.1$~GPa  is the appearance of three
new Raman modes at 201, 290 and 363 cm$^{-1}$. The original five Raman modes at
ambient pressure change slightly at $p_1$. According the group theory
and the first-principle calculations of the single $\alpha$-RuCl$_3$ layer, we
assign the mode at 201~cm$^{-1}$ as an IR-active $E_u$ mode, 290~cm$^{-1}$  as an inactive
$A_{2g}$ mode, and 363 cm$^{-1}$ as an IR-active $A_{2u}$  mode
which has the highest frequency and is related to the asymmetrical layer breathing~\cite{Note1}. The appearance of IR-active modes in Raman
spectrum indicates the inversion symmetry breaking.  To confirm this, the second harmonic generation (SHG)
measurement is performed on pressured samples~\cite{Note1}.
No SHG signal was detected before $p_1=1.1$~GPa, but SHG signal was detected at 1.4 GPa
and higher pressure, which is consistent with inversion symmetry breaking at
around $p_1=1.1$~GPa. Because of no significant change in other Raman modes, we can
conclude that the inversion symmetry breaking is due to stacking pattern change
of the $\alpha$-RuCl$_3$ layers.

\textit{Dimerization transition at $p_2=1.7$~GPa --}
As shown in Fig.~\ref{fig:figure2} (b), almost all of the Raman modes show blue-shift with
the pressure increasing, 
however, the Ru in-plane mode at 161~cm$^{-1}$ displays anomalous red-shift and a large splitting at $p_2=1.7$~GPa. As shown in Fig.~\ref{fig:figure2} (b), a
split peak appears at around 171 cm$^{-1}$ after 1.7~GPa and rapidly increases
to 181 cm$^{-1}$ at around 2.4~GPa. The  the split peak has the higher frequency indicating
that two Ru atoms move close to each other and form the dimerization state. By simply assuming that the distance
  of the nearest Ru atoms is inversely proportional with the frequency of the Ru
  in-plane mode, we can estimate that the Ru-Ru bond dimerization is about
  0.5~\AA   ~at 2.4~GPa and increases with pressure.


The Ru-Ru dimerization in the $\alpha$-RuCl$_3$ layers splits all degenerated
$E$ modes into $A + B$ modes. We do observe such a splitting for the
twist mode at 116~cm$^{-1}$, shearing mode at 268~cm$^{-1}$ (it splits at lower pressure probably due to the
interlayer interaction). We don't detect the splitting
for the ring breathing mode at 294~cm$^{-1}$ probably due to its low intensity and the adjacent
intensive layer breathing mode.
By considering the normal modes of the split Ru
in-pane mode, we can assign the high frequency one as $A_g$ and the low
frequency one as $B_g$. As shown in Fig.~\ref{fig:figure1}, the $A_g$ peak becomes much more
intense than the $B_g$ peak with pressure increasing. By using 488~nm laser as
excitation light, the $B_g$ peak even becomes unresolvable~\cite{Note1}. Similarly, the $B_g$ peak of the split twist mode is the low
frequency one and cannot be observed after the dimerization. For the shearing
mode, the $B_g$ peak is the high frequency one and disappears after the phase
transition. In a summary, the softening and large splitting of the Ru in-plane
mode provide a direct evidence of dimerization and the behaviors of other Raman
modes are consistent with this picture.

\begin{figure}[t]
\centering
\includegraphics[width=\columnwidth]{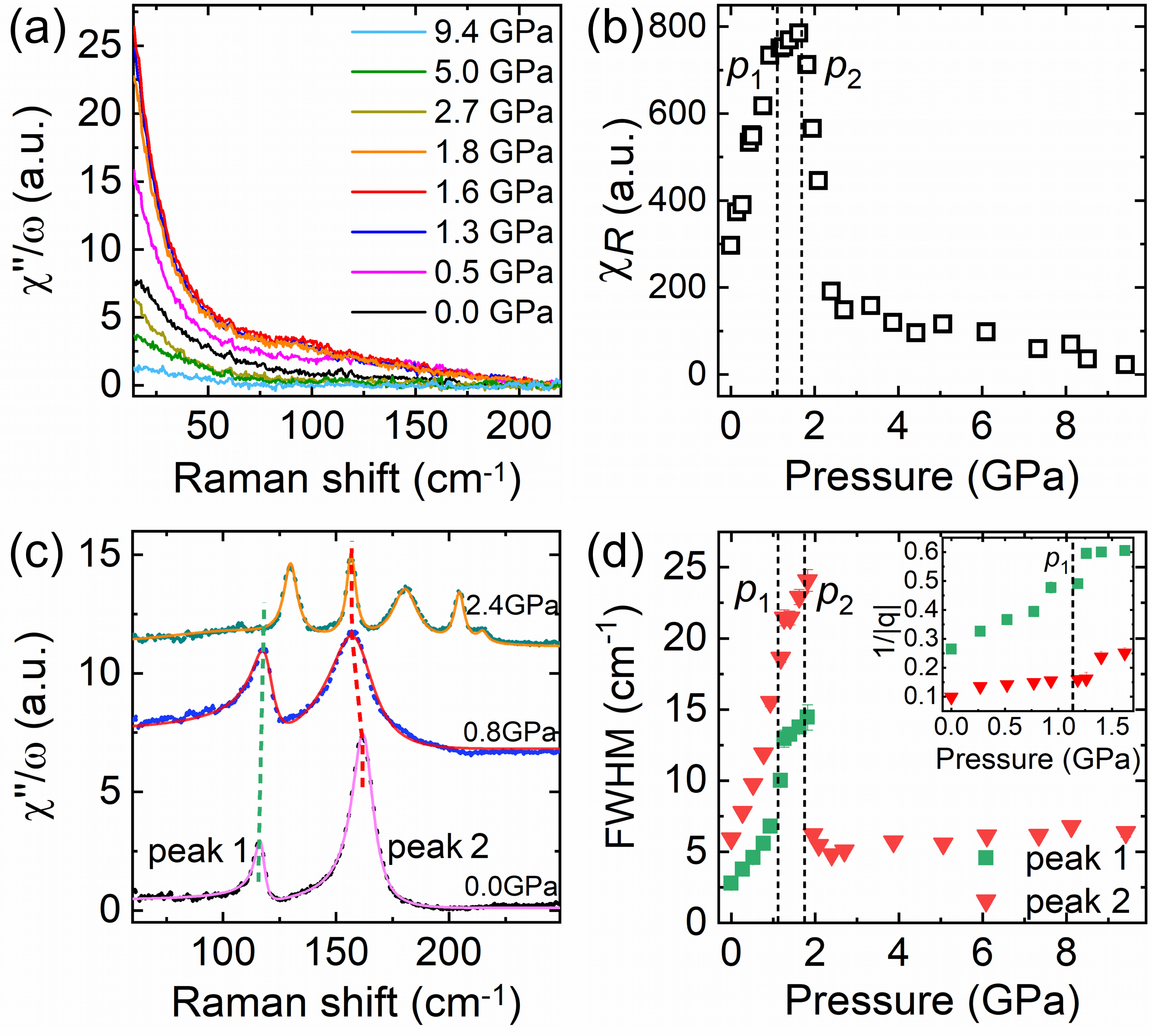}
\caption{ Pressure dependence of the magnetic Raman conductivity
  $\chi''/\omega$ (a) and the magnetic Raman susceptibility
  $\chi_R$ (b).  (c) Data and fittings of the phonon Raman spectra (after subtracting the magnetic continuum) for the twisting mode (peak 1) and the Ru in-plane mode (peak 2) under various pressures. The spectra under 0.0~GPa and 0.8~GPa were fitted by Fano peaks. The spectrum under 2.4~GPa was fitted by Lorentz peaks.
  (d) Pressure dependence of the full width at half maximum (FWHM) and the Fano asymmetry parameter $1/|q|$ (the inset) for the peak 1 and peak 2. 
  }
\label{fig:figure3}
\end{figure}

\textit{Magnetic breakdown from Raman susceptibility --}
Raman spectroscopy also measures the magnetic response in the
strong spin-orbit coupling system~\cite{Moriya1968,Fleury1968,Lemmens2003,Nasu2016}. The Raman intensity $I(\omega)$ is
proportional to the dynamical Raman tensor susceptibility, $I(\omega) \propto
[1+n(\omega)]\chi''(\omega)$. Here $\chi''(\omega)$ is the imaginary part of
the correlation functions of Raman tensor, $\chi(\omega)=\int_{0}^{\infty}dt\int
d\mathbf{r}\langle\{\tau(0,0),\tau(\mathbf{r},t)\}  \rangle e^{-i\omega t}$. In general, we can expand the Raman tensor
$\tau(\mathbf{r})$ in powers of spin-1/2 operators, $\tau^{\alpha\beta}(\mathbf{r})=\tau_0^{\alpha\beta}(\mathbf{r})+\sum_\mu
K^{\alpha\beta}_{\mu}S^\mu(\mathbf{r})+\sum_{\delta}\sum_{\mu\nu}M^{\alpha\beta}_{\mu\nu}(\mathbf{r},\delta)S_{\mathbf{r}}^\mu
S_{\mathbf{r}+\delta}^\nu+...$. The first term corresponds to Rayleigh
scattering, the second and third term are linear and quadratic in the spin
operators $S_{\mathbf{r}}^\alpha$ and correspond to the one magnon~\cite{Moriya1968,Fleury1968,Lemmens2003} and two-magnon process~\cite{Moriya1968,Fleury1968,Lemmens2003,Knolle2014a,Nasu2016,Fu2017}, respectively. The
complex tensor $K$ determines the strength of the coupling of light to
the spin system associated with spin-orbit coupling.

\begin{figure}[b]
\centering
\includegraphics[width=0.9\columnwidth]{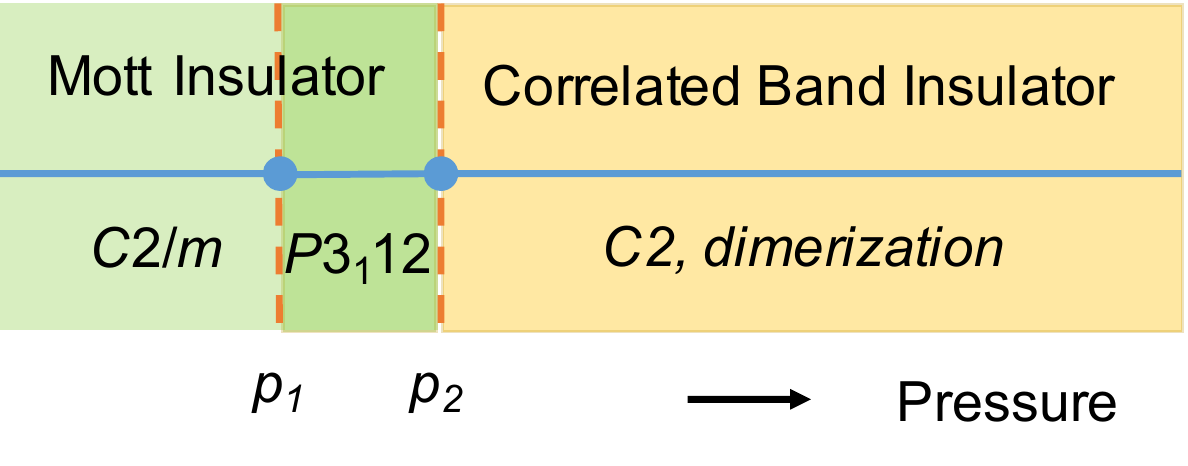}
\caption{Schematic phase diagram of $\alpha$-RuCl$_3$ under pressures. The crystal structure changes from $C2/m$ to $P3_112$ at $p_1$, and further to $C2$ at $p_2$ due to the Ru-Ru bond dimerization. Meanwhile, before the dimerization at $p_2$, the system is a Mott insulator with magnetism; after dimerization, it is a correlated band insulator without magnetism.}
\label{fig:figure4}
\end{figure}
To extract the Raman susceptibility, we
first get Raman tensor conductivity $\chi''(\omega)/\omega$ for frequencies down to $15$
cm$^{-1}$, as shown in Fig.~\ref{fig:figure3} (a) and then integrated over the
frequency rang of 15-220~cm$^{-1}$ to get the Raman susceptibility
$\chi_R\equiv\frac{2}{\pi}\int\frac{\chi''(\omega)}{\omega}d\omega$ in Fig.~\ref{fig:figure3} (b),  by
using the Kramers-Kronig relation.   The pressure
dependent of $\chi_R$ manifests a rapid increase with increasing the
pressure and then a sharp drop to zero at $p_2=1.7$~GPa. The Raman susceptibility $\chi_R$ contains
the static spin susceptibility and multi-spin susceptibility (\textit{e.g.},
bond spin operators), corresponding to one-magnon and multi-magnon process, respectively. 
As we notice that the static magnetic susceptibility $\chi_m$ of
$\alpha$-RuCl$_3$ at room temperatures changes little reported in
Refs.~\cite{Cui2017,Biesner2018} before dimerization, we suspect that
the increasing Raman susceptibility $\chi_R$ mainly comes from the multi-magnon
processes. Above $p_2=1.7$~GPa, both magnetic susceptibility $\chi_m$ and Raman
susceptibility $\chi_R$ break down, implying the dimerized non-magnetic state.

 The Raman spectra of the phonon modes at 116~cm$^{-1}$ and
161~cm$^{-1}$ display significant Fano asymmetry before $p_2=1.7$~GPa as shown
in Fig.~\ref{fig:figure3} (c). It captures the mutual couplings between the lattice and magnetic
excitations.  
The full width at half maximum (FWHM) and the Fano asymmetry parameter $1/|q|$ measure the strength of the coupling between
the lattice and magnetic excitations, as shown in Fig.~\ref{fig:figure3} (d). We
can see that the coupling between lattice and magnetic excitations increases with pressure and is
completely suppressed after dimerization, consistent with the evolution of Raman susceptibility
$\chi_R$ in Fig.~\ref{fig:figure3} (b).

\textit{Discussions and conclusions--}
$\alpha$-RuCl$_3$ has a monoclinic $C2/m$ structure at room temperature at ambient
pressure. At $p_1=1.1$~GPa, the inversion symmetry is breaking and the system probably turns
into the trigonal $P3_112$ structure where the inter-layer interaction distorts
the inversion symmetry. The
structural transition at $p_1=1.1$~GPa is first-order type since the space group changes.  The softening and
big splitting of the Ru in-plane mode at $p_2=1.7$~GPa provide a
direct evidence of dimerization. However, the softening is not complete, but a
little. According the ``little phonon softening'' theory~\cite{Krumhansl1989},
the structural dimerization at $p_2=1.7$~GPa is a first-order transition. We suspect the system has the space group $C2$ after the dimerization.
 Dimerization brings the magnetic breakdown due to Mott
collapse and the system remains insulating after dimerization. The schematic phase
diagram is summarized in Fig.~\ref{fig:figure4}.

More remarks on the Mott collapse are needed here. At ambient pressure, $\alpha$-RuCl$_3$ is a spin-orbit Mott insulator with the
Kitaev magnetism. Comparing to the iridates,  $\alpha$-RuCl$_3$ has a larger
electron correlation, but the effective $j_{\text{eff}}=1/2$ and $3/2$ bands
near the Fermi surface are not well separated due to a smaller spin-orbit
coupling. The mixing between the effective $j_{\text{eff}}=1/2$ and $3/2$ bands  brings $\alpha$-RuCl$_3$ closer to a
Mott transition. According to the first-principle calculations~\cite{Kim2015}, when
the on-site Coulomb interaction $U$ is introduced while fixing a paramagnetic
state, the bands near the Fermi level take on a predominantly
$j_{\text{eff}}=1/2$ character and a band gap develops, suggesting a
correlation-induced insulating phase. The pressure increases the band width, and
hence the mixing between the effective $j_{\text{eff}}=1/2$ and $3/2$ bands. It
would finally drive the system into a correlated band insulator via the Mott
collapse. The Ru-Ru bond dimerization accelerates the Mott collapse process. Magnetism and chemical bondings is also studied the isostructural counterpart $\alpha$-MoCl$_3$ by the Raman scattering~\cite{McGuire2017}. The spin-orbit coupling and quantum spin liquid physics brings more significant relativity and quantum effects in the studies of magnetism and chemical bonds~\cite{Goodenough1963}.

In conclusions, we perform Raman studies on the relation between Kitaev
magnetism and chemical bondings in $\alpha$-RuCl$_3$ under pressures. At the critical
pressure $p_1=1.1$~GPa, $\alpha$-RuCl$_3$ undergoes the structural transition
with the inversion symmetry breaking from the monoclinic $C2/m$ to the trigonal
$P3_112$. due to different layer stacking. At the critical $p_2=1.7$~GPa, Ru-Ru
bonds in the $\alpha$-RuCl$_3$ dimerizes, and the system turns out to a
correlated band insulator due to the Mott collapse.

\textit{Note added --} During the preparation of our manuscript, similar results about the structural phase transition studied by XRD and infrared spectroscopy were reported by other researchers\cite{Bastien2018,Biesner2018}.

\acknowledgements{\textit{Acknowledgments --}
  We would like to thank Prof. Hugen Yan for the IR measurement on the
  exfoliated $\alpha$-RuCl$_3$ layers. J.W.~Mei thanks Dr. Shunhong Zhang for
  useful discussions at the early stage of this work. This work was supported by the Science,Technology and Innovation Commission of Shenzhen Municipality (Grant No.ZDSYS20170303165926217). M.H. was partially supported by the Science,Technology and Innovation Commission of Shenzhen Municipality (Grant No.JCYJ20170412152334605 and JCYJ20160531190446212).  F.Y. was partially supported by National Nature Science Foundation of China 11774143 and JCYJ20160531190535310. }

\bibliography{rucl3}

\end{document}